\documentclass[10pt,aps,prl,twocolumn,superscriptaddress,showkeys,nobibnotes,secnumarabic]{revtex4-1}
\usepackage{graphicx}
\usepackage{longtable}
\usepackage{color}
\usepackage{changes}
\usepackage{amsmath}
\usepackage{upgreek}
\newcommand{\cm}{cm$^{-1}$\,}
\begin{document}

\title{Carbon nanotube chirality determines properties of encapsulated linear carbon chain}%

\author{Sebastian Heeg}%
\affiliation{ETH Z\"urich, Photonics Laboratory, 8093 Z\"urich, Switzerland}
\author{Lei Shi}
\affiliation{University of Vienna, Faculty of Physics, 1090 Wien, Austria}
\author{Lisa V. Poulikakos}
\affiliation{ETH Z\"urich, Optical Materials Engineering Laboratory, 8093 Z\"urich, Switzerland}
\author{Thomas Pichler}
\affiliation{University of Vienna, Faculty of Physics, 1090 Wien, Austria}
\author{Lukas Novotny}
\email{lnovotny@ethz.ch}
\homepage{www.photonics.ethz.ch}
\affiliation{ETH Z\"urich, Photonics Laboratory, 8093 Z\"urich, Switzerland}


\begin{abstract}
Long linear carbon chains encapsulated inside carbon nanotubes are a very close realization of carbyne, the truly one-dimensional allotrope of carbon. Here we study individual pairs of double-walled carbon nanotubes and encapsulated linear carbon chains by tip-enhanced Raman scattering. We observe that the radial breathing mode of the inner nanotube correlates with the frequency of the carbon chain's Raman mode, revealing that the nanotube chirality determines the vibronic and electronic properties of the encapsulated carbon chain. We provide the missing link that connects the properties of the encapsulated long linear carbon chain with the structure of the host nanotube. 
\end{abstract}

\maketitle
%
%
Carbyne by definition is an infinitely long linear carbon chain with $\mathit{sp}^{1}$ hybridization that forms the truly one-dimensional allotrope of carbon at the one-atom cross section limit~\cite{Heimann:2012ig}. Its anticipated stiffness, strength, and elastic modulus exceed that of any other known material~\cite{Liu:2013km}. In its most common form, carbyne is a polyyne with alternating single and triple bonds originating from a Peierls distortion~\cite{Peierls:1955jy}. This bond-length alternation (BLA) dominates the electronic and vibronic structure of carbyne~\cite{Kurti:1995hp,Milani:2008fwa,Yang:2006fq}. It opens up a direct band gap that is sensitive to external pertubations thus offering tunability. Finite linear carbon chains are expected to exhibit the properties of carbyne if they consist of $100$ or more atoms as the BLA saturates~\cite{Heimann:2012ig}. Exploring the fundamental properties of carbyne experimentally, however, has long been hindered by its extreme chemical instability and short chain lengths of up to $44$ atoms~\cite{Chalifoux:2010fp}.  

\begin{figure}[t]
\includegraphics[width=\linewidth]{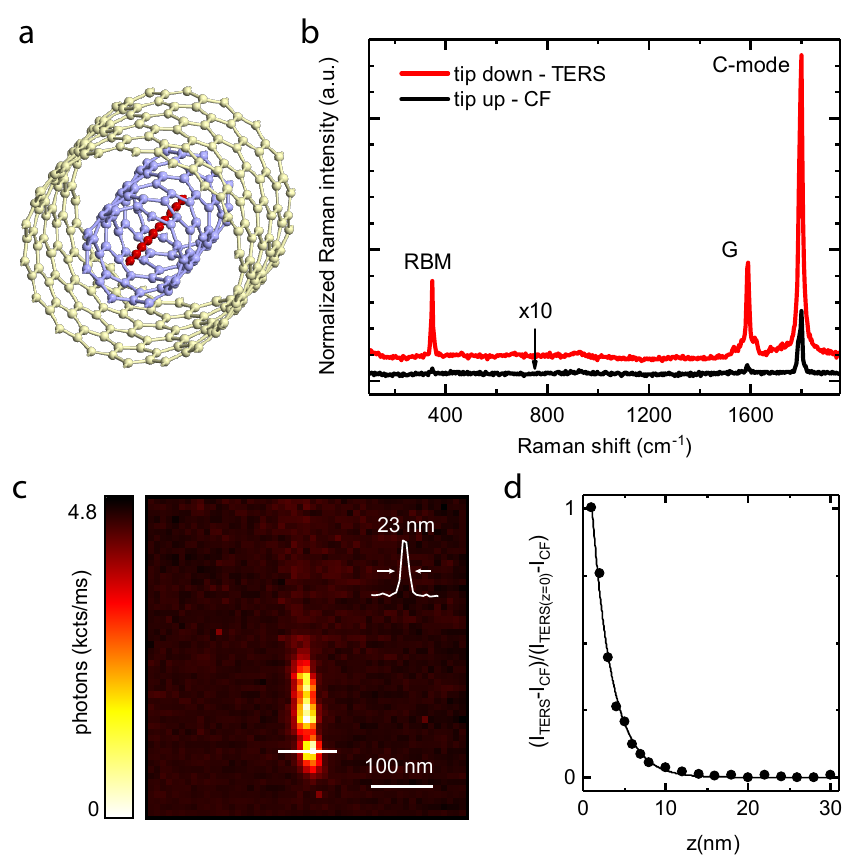}
  \caption{TERS characterization of long linear carbon chains. (a) Sketch of a linear carbon chain (red) encapsulated in a double-walled carbon nanotube (blue/yellow). (b) Confocal (black) and TERS (red) Raman spectra of the same double-walled carbon nanotube and encapsulated long linear carbon chain. (c) TERS image of a linear carbon chain. The image contrast corresponds to the intensity of the C-mode centered at $1798\,$cm$^{-1}$. An intensity profile (inset) extracted along the white line shows a resolution of $\sim23\,$nm. (d) Normalized TERS C-mode intensity as function of the separation $z$ between tip and sample. The solid line is an exponential fit with a decay length of $2.5\,$nm.}
  \label{FIG:TERS}
 \end{figure}

The synthesis of linear carbon chains inside carbon nanotubes, sketched in Fig.~\ref{FIG:TERS}(a), overcomes these obstacles~\cite{Zhao:2003dz, Fantini:2006ix,Jinno:2006gm,Nishide:2007bd,Shi:2011ey,Andrade:2015fka,Andrade:2015ke}. The tubes act as nano-reactors, prevent chemical interaction of the chains with the environment, and allow for long chain lengths. A major leap forward in forming carbyne are long linear carbon chains (LLCCs) with lengths up to several hundreds of nanometers that have recently been synthesized inside double-walled carbon nanotubes (DWCNTs)~\cite{Shi:2016kqa,Shi:2017cd,Lapin:2015bo,Rohringer:2016jx}. The local environment inside a nanotube, characterized by its chirality, strongly affects encapsulated chains through interactions such as van-der-Waals (vdW) forces, charge transfer or dielectric screening, and may even limit the chain length~\cite{Zhao:2003dz, Fantini:2006ix,Jinno:2006gm,Nishide:2007bd,Shi:2011ey,Andrade:2015fka,Andrade:2015ke,Shi:2016kqa,Rohringer:2016jx,Shi:2017cd,Lapin:2015bo,Wanko:2016kq}. These interactions vary with chirality, modify the BLA, and effectively mask the intrinsic properties of LLCCs~\cite{Wanko:2016kq}. However, while the properties of the chains are governed by its CNT host, the correspondence between the tube's chirality and the properties of the encapsulated chain has not been investigated experimentally.

Raman spectroscopy is an excellent tool to study the properties of the encapsulated chain and the characteristics of the encasing CNT. The dominant Raman mode of carbyne (C-mode) reports the bond-length alternation of the chain~\cite{Heimann:2012ig}. The chirality specific Raman features of CNTs, in particular the radial breathing mode (RBM), are very well understood~\cite{Reich:2009wo,Maultzsch:2005fp}. Accordingly, Raman measurements of nanotubes hosting LLCCs allow us to correlate the electronic and vibronic properties of the LLCC to the properties of the nanotube and help to identify the dominating interaction between host and guest. Importantly, such a correlation cannot be established through bulk measurements because it is impossible to verify that the Raman signal of both chain and tube arises from the same pair. 

Tip-enhanced Raman scattering (TERS), on the other hand, offers the nanoscale spatial resolution necessary to find and characterize individual pairs of carbon nanotube and chain, measure the length of the chain, and unveil the correlation between the nanotube's chirality and the chain's properties~\cite{Shi:2016kqa,Shi:2017cd,Lapin:2015bo}. In this letter, we present TERS measurements that quantitatively link the C-mode frequency of long linear carbon chains to the chirality of the encapsulating carbon nanotubes. We observe that the frequency of the C-mode decreases as the inner nanotube diameter is reduced.

Details of our experimental setup have been described in previous reports~\cite{Hartschuh:2003co,Hartschuh:2008eea,Johnson:2012iq}. In short, a focused radially-polarized laser beam ($\lambda=633\,$nm) irradiates a gold pyramid fabricated by template-stripping~\cite{Johnson:2012iq}. The enhanced field at the tip apex defines a localized excitation source for Raman scattering. Raman scattered light is collected in backscattering configuration either by a combination of bandpass filters that transmit the spectral region of the linear carbon chain's C-mode ($1770\,$cm$^{-1}$ to $1870\,$cm$^{-1}$~\cite{Zhao:2003dz, Fantini:2006ix,Jinno:2006gm,Nishide:2007bd,Shi:2011ey,Andrade:2015fka,Shi:2016kqa,Andrade:2015ke,Rohringer:2016jx,Shi:2017cd,Lapin:2015bo}) followed by a single-photon counting avalanche detector, or by a combination of spectrograph and charge-coupled device (resolution $\sim3\,$cm$^{-1}$). To avoid sample degradation due to heating, we used laser powers of $115\,\upmu$W for TERS and $1.15\,$mW for confocal (CF) imaging. Integration times are $25\,$ms/pixel for imaging and up to $60\,$s for acquiring full Raman spectra. LLCCs were grown inside double-walled carbon nanotubes (DWCNTs) in a high-temperature and high vacuum process as described in Ref.~\cite{Shi:2016kqa}. The tubes were then purified, individualized, and dispersed on thin glass cover slides using chlorosulfonic acid in an oxygen free atmosphere. 

To locate individual carbon chains inside DWCNTs we first perform confocal Raman raster scans (not shown). Once the signature of a linear carbon chain is detected, we verify its presence by observing the C-mode of the chain in a confocal Raman spectrum (black) as shown in Fig.~\ref{FIG:TERS}(b). We then position the tip near the surface ($<1\,$nm) and record high resolution topographical (see Supporting Information)  and TERS images as shown in Fig.~\ref{FIG:TERS}(c). The TERS image reveals that the DWCNT contains a linear carbon chain of length $200\,$nm. The inset in Fig.~\ref{FIG:TERS}(c) shows a line profile (white) extracted from the TERS image indicating a resolution of $23\,$nm.
 
The TERS spectrum (red), shown in Fig.~\ref{FIG:TERS}(b), exhibits a peak at $1798\,$cm$^{-1}$ originating from the chain's C-mode, a longitudinal-optical phonon that arises from the in-phase stretching of the triple-bonds along the chain. It is the only Raman-active phonon in polyynic carbyne. The group of peaks with the dominant component at $1588\,$cm$^{-1}$ are the G-modes, typical signatures of CNTs~\cite{Reich:2009wo}. The strong RBM at $348\,$cm$^{-1}$ belongs to the inner carbon nanotube that is in resonance with the excitation wavelength~\cite{Pfeiffer:2005ho,Anonymous:XSF3TiDN}. The RBM corresponds to a radial displacement of the carbon atoms forming the nanotube and enables us to determine its chirality~\cite{Reich:2009wo,Maultzsch:2005fp,Thomsen:2007vc}. We do not observe any signature of the defect-induced D-mode ($\sim 1350\,$cm$^{-1}$), indicating that the nanotubes are pristine and largely free of defects~\cite{Reich:2009wo}. Upon increasing the tip-sample distance the Raman intensity drops exponentially as shown for the C-mode in Fig.~\ref{FIG:TERS}(d), which confirms the local and enhanced nature of our TERS signal as compared to the confocal Raman spectra (black) in Fig.~\ref{FIG:TERS}(b).

 \begin{figure}[t]
\includegraphics[width=\linewidth]{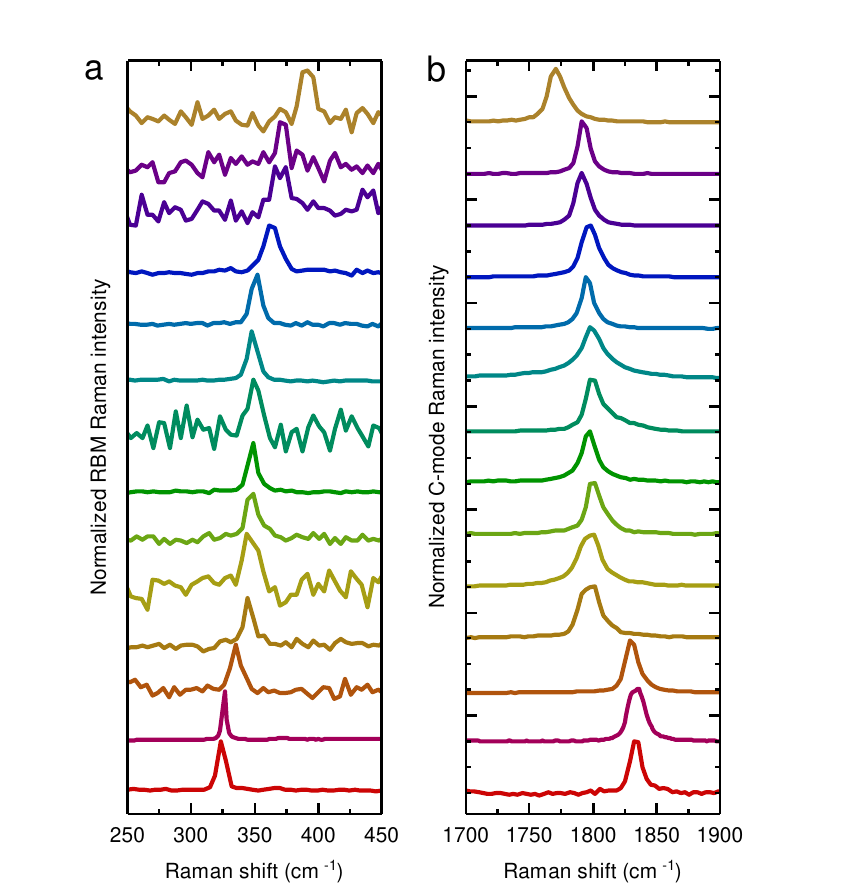}
  \caption{TERS spectra of 14 pairs of long linear carbon chains and encapsulating carbon nanotubes. (a)~RBM spectra of the inner nanotube from highest (top) to lowest frequency. (b) C-mode Raman modes of the chains corresponding to the RBMs shown in (a). RBM and C-mode spectra are normalized independently to the same peak height.}
  \label{FIG:RBM_LLCC_SPECTRA}
 \end{figure}

We identified and characterized $14$ different chain-nanotube systems that exhibit both a single C-mode frequency and a single RBM frequency. We present the corresponding TERS spectra with a focus on RBMs in Fig.~\ref{FIG:RBM_LLCC_SPECTRA}(a) and C-modes in Figs.~\ref{FIG:RBM_LLCC_SPECTRA}(b). As we are interested in peak positions only, the RBMs and C-modes are independently normalized to the same amplitude and offset for clarity with decreasing RBM frequency from top to bottom. For all spectra in Fig.~\ref{FIG:RBM_LLCC_SPECTRA}, a single RBM can clearly be assigned to a corresponding single C-mode. 

\begin{figure}[t]
\includegraphics[width=\linewidth]{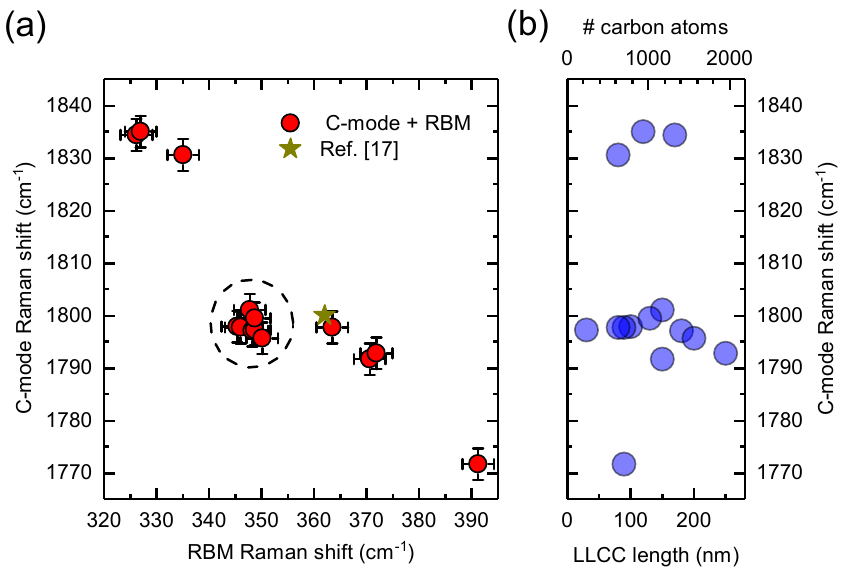}
  \caption{Correlation between C-mode and RBM. (a) C-mode of encapsulated linear carbon chains vs radial breathing mode of the encasing inner carbon nanotubes extracted from the TERS spectra shown in Fig.~\ref{FIG:RBM_LLCC_SPECTRA}. (b) Length of LLCCs shown in (a) as observed by TERS in nm and as number of carbon atoms \# forming the LLCC with $0.2558\,$nm lattice constant~\cite{Shi:2016kqa}.}
  \label{FIG:RBM_LLCC_CHIRALITY}
 \end{figure}

We now analyze in detail the correlation between chain and encasing nanotube by extracting the peak positions from Fig.~\ref{FIG:RBM_LLCC_SPECTRA} and plotting the C-mode frequency as a function of RBM frequency in Fig.~\ref{FIG:RBM_LLCC_CHIRALITY}(a), including the singular data point from Ref.~\cite{Lapin:2015bo}. The plot reveals three distinct characteristics. First, each RBM -- representing the chirality of the encasing nanotube -- uniquely corresponds to one specific C-mode frequency -- representing the linear chain -- to within $\pm\,3\,$cm$^{-1}$. We refer to such a combination of RBM and C-mode from now on as 'pairing'. Second, specific pairings of RBM and C-mode occur multiple times. The pairing for the RBM at $348\,$cm$^{-1}$ and the C-mode at $1798\,$cm$^{-1}$, dashed circle in Fig.~\ref{FIG:RBM_LLCC_CHIRALITY}(a), is particularly pronounced and applies to half of our measurements. These RBMs arise from the same inner nanotube chirality and the encapsulated chains show similar Raman frequencies. The length of the chains, depicted in Fig.~\ref{FIG:RBM_LLCC_CHIRALITY}(b) for all measured pairings, does not affect the chain's C-mode frequencies. This is the core observation of our experiments because it is in stark contrast to the pronounced length dependence of the C-mode frequencies for short linear carbon chains. Our measurements reveal a unique correspondence between nanotube RBM and carbon chain C-mode frequencies, independent of the length of the carbon chain as observed by TERS. We conclude that the inner carbon nanotube chirality represented by this RBM uniquely determines the chain's bond-length alternation and hence its vibronic and electronic structure. 

 \begin{table}[b]
\begin{ruledtabular}
\begin{tabular}{l | c c | c c | c | c c}
RBM (cm$^{-1}$) & $327\pm1$ & $335$ & $348\pm2$ & $364$ & $371\pm1$ & $391$ \\
(n,m) CNT$_{\textrm{in}}$ & \multicolumn{2}{c|}{(6,5)}  & \multicolumn{2}{c|}{(6,4)}  & (7,2) & (8,0) \\
d (nm) & \multicolumn{2}{c|}{0.747} & \multicolumn{2}{c|}{0.683} & 0.641 & 0.626 \\
\hline
C-m. (cm$^{-1}$) & $1835\pm1$ & $1831$ & $1798\pm2$ & $1798$ &  $1792\pm1$ & $1772$ \\
E$_{\textrm{g}}$(eV)~\cite{Shi:2017cd} & 2.08 & 2.05 & \multicolumn{2}{c|}{1.86} & 1.84 & 1.76 \\

\end{tabular}
\end{ruledtabular}
\caption{\label{TAB:CHIRALITIES} RBM frequencies, chiralities, and diameters of the inner nanotubes together with the associated C-mode frequencies and band gaps of the encapsulated chains. The diameter $d$ of a $(n,m)$ nanotube is given by $d=a/\pi \sqrt{n^2 + nm+m^2}$ with $a=0.246\,$nm~\cite{Reich:2009wo}.}
\end{table}  

Before we assign RBMs to nanotube chiralities, we verify that the presence of a linear carbon chain does not modify the RBMs of inner and outer nanotubes, which could otherwise falsify our chirality assignment. We show this for one of the pairings ($326\,$cm$^{-1}$/$1835\,$cm$^{-1}$), where we also observe the RBM of the outer nanotube at $187\,$cm$^{-1}$. This pairing is compared to an empty DWCNT formed by the same inner and outer nanotube chiralities, see Supporting Information. For both the empty and the filled DWCNT we observed the same RBM frequencies and conclude that the RBMs are generally not affected by the presence of encapsulated LLCCs. 

Assigning RBMs to chiralities is not straightforward for double-walled carbon nanotubes, because one particular inner tube can reside in different outer tubes. This variation in the diameter difference between inner and outer tube influences the wall-to-wall interactions. It renders the simple inverse relation between nanotube diameter and RBM frequency for single-walled carbon nanotubes invalid~\cite{Reich:2009wo,Thomsen:2007vc}. Instead, the inner tube's RBM shifts up in frequency depending on the outer nanotube's diameter, while the diameters of both inner and outer tubes effectively remain unchanged~\cite{Pfeiffer:2005ho,Anonymous:XSF3TiDN,Popov:2005jx,Dobardzic:2003ct}. As a consequence, DWCNT spectra show more RBMs than there are chiralities and RBMs of different frequencies may belong to the same inner tube chirality. Table~\ref{TAB:CHIRALITIES} lists our tentative assignment of the pairings of RBM and C-mode in Fig.~\ref{FIG:RBM_LLCC_CHIRALITY}(a) to the chiralities of the inner carbon nanotubes and the corresponding tube diameters $d$. The assignment is based on experimental Raman studies on DWCNTs and described in detail in the Supporting Information. 

We find that the two RBMs around $327\,$cm$^{-1}$ and $335\,$cm$^{-1}$ belong to the $(6,5)$ inner nanotube, while the RBMs around $348\,$cm$^{-1}$ and $364\,$cm$^{-1}$ belong to the $(6,4)$ inner nanotube. These assignments intuitively make sense because chains with similar C-modes are allocated to inner tubes of the same chirality. The variation in the C-mode frequency for different pairings belonging to the same inner nanotube is then a measure for the effect of the outer nanotube on the encapsulated chain. We find this effect to be comparably small with a maximal variation of $5\,$cm$^{-1}$ for the $(6,5)$ nanotube, c.f. Table~\ref{TAB:CHIRALITIES}. 

\begin{figure}[t]
\includegraphics[width=\linewidth]{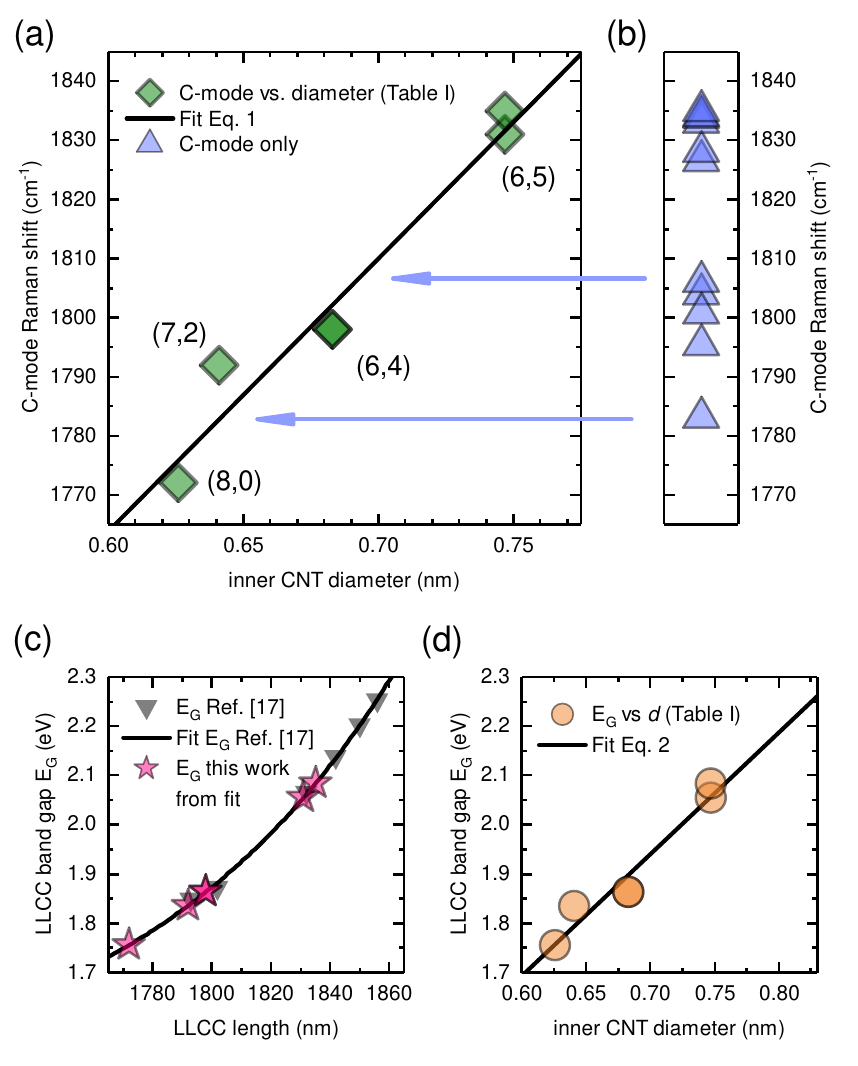}
  \caption{Correlation between diameter of encasing inner carbon nanotube vs C-mode frequency and bandgap of encapsulated long linear carbon chain. (a) C-mode frequency of LLCC as function of the encasing inner tube's diameter. (b) C-mode frequencies of LLCCs for which no RBM from the inner tube could be detected. (c) Band gap $E_G$ as a function of C-mode from Ref.~\cite{Shi:2017cd} (triangles) together with an exponential fit that provides $E_G$ for the C-modes in this work (stars). (d) LLCC Band gap as function of inner CNT diameter.}
  \label{FIG:DIAM_BGS}
 \end{figure}

Our third observation from the plot in Fig.~\ref{FIG:RBM_LLCC_CHIRALITY}(a) is the decrease of the C-mode frequency with increasing RBM frequency. Making use of our assignment of RBMs to inner nanotube chiralities, c.f. Table~\ref{TAB:CHIRALITIES}, we plot in Fig.~\ref{FIG:DIAM_BGS}(a) the C-mode frequencies of the encapsulated chains as a function of the encasing inner carbon nanotube's diameter. The data shows a clear decrease in the chain's C-mode frequency $\omega_{\rm{C}}$ with decreasing diameter $d$ and yields the linear relation 
\begin{equation}\label{EQ:CMODE_DIAMETER} 
\omega_{\rm{C}}(d)=a + b\, \cdot d,
\end{equation}
with $a=1487\pm36\,$cm$^{-1}$ and $b=462\pm41\,$cm$^{-1}$nm$^{-1}$. This relation excellently describes our observations and is plotted as the black line in Fig.~\ref{FIG:DIAM_BGS}(a). Note that we use the average C-mode frequency given in Table~\ref{TAB:CHIRALITIES} instead of all $14$ measured C-mode values for fitting Eq.~\ref{EQ:CMODE_DIAMETER} such that each pairing carries equal weight. 

The linear decrease of the encapsulated chain's C-mode frequency with decreasing nanotube diameter is strong evidence that the properties of the encapsulated chains are dominated by vdW interactions as this trend has recently been postulated by \textit{Wanko et al.}~\cite{Wanko:2016kq}. By measuring chains encapsulated in different DWCNTs, our TERS measurements randomly probe different levels of charge transfer that vary with the four different inner CNT chiralities, labeled in Fig.~\ref{FIG:DIAM_BGS}(a), that are in resonance with the excitation. However, we find the effect of charge transfer on C-mode frequency to be small and a possible cause for the deviations from Eq.~\ref{EQ:CMODE_DIAMETER}, which amounts to maximally $9\,$cm$^{-1}$ for the $(7,2)$ inner nanotube.  

Equation~\ref{EQ:CMODE_DIAMETER} provides a robust estimate of the inner nanotube's diameters based on the C-mode frequency alone. This estimate is relevant for most of our $\sim75$ TERS measurements, as only a few of the inner nanotube chiralities present in the sample are in resonance with our excitation wavelength, c.f. Fig.~\ref{FIG:RBM_LLCC_CHIRALITY}(a)  and Table~\ref{TAB:CHIRALITIES}, and the resonance window of the RBM only spans around $50\,$meV~\cite{Reich:2009wo,Maultzsch:2005fp,Thomsen:2007vc}. Very rarely this is compensated by TERS enhancement. Figure~\ref{FIG:DIAM_BGS}(b) lists a representative set of measured C-mode frequencies for which no RBM appeared in the corresponding TERS spectrum. The blue arrows illustrate the assignment to diameters following Eq.~\ref{EQ:CMODE_DIAMETER}. We expect the C-mode at $1808\,$cm$^{-1}$, for instance, to reside inside a tube with $d=0.694\,$nm, very close to the $(7,3)$ chirality with $d=0.696\,$nm, the resonance of which is at $2.5\,$eV and hence far from our excitation ($1.96\,$eV).

We now compare our observations to recent wavelength-dependent Raman measurements on dense samples of LLCCs in DWCNTS~\cite{Shi:2017cd}. In these bulk measurements, the Raman spectra were dominated by six C-mode frequency bands at $1793\,$cm$^{-1}$ to $1856\,$cm$^{-1}$. Three of these bands are within our measurement range ($1793\,$cm$^{-1}$, $1802\,$cm$^{-1}$, and $1832\,$cm$^{-1}$) and agree well with data, c.f. Fig~\ref{FIG:RBM_LLCC_CHIRALITY}(a) and (b), where we also find a strong clustering of the C-modes around $1800\,$cm$^{-1}$ and $1830\,$cm$^{-1}$. It suggests that Eq.~\ref{EQ:CMODE_DIAMETER} connecting the diameter of the inner carbon nanotube to the C-mode frequency also holds for higher C-mode frequencies not observed in our TERS experiments. The highest observed C-mode at $1856\,$cm$^{-1}$ corresponds to an inner tube diameter of $d=0.80\,$nm, well in the diameter range required for the growth of long linear carbon chains. 

Equation~\ref{EQ:CMODE_DIAMETER} allows us to express the band gaps $E_G$ of encapsulated linear carbon chains as a function of the encasing inner nanotube's diameter. We extract $E_G$ corresponding to a particular C-mode frequency from Ref.~\cite{Shi:2017cd} and plot the data (triangles) in Fig.~\ref{FIG:DIAM_BGS}(c) together with the best fit which we find to be of exponential form, see Supporting Information. This allows us to obtain $E_G$ corresponding the C-modes observed here, stars in Fig.~\ref{FIG:DIAM_BGS}(c) and Table~\ref{TAB:CHIRALITIES}, and to plot $E_G$ as a function of the inner nanotube's diameter in Fig.~\ref{FIG:DIAM_BGS}(d). We find the band gap of encapsulated chains to increase linearly with the diameter of the inner nanotube as 
\begin{equation}\label{EQN:BG_DIAM}
E_G(d)= c+f\cdot d,
\end{equation}
with $c=0.21\pm0.2\,$eV and $f=2.47\pm0.3\,$eV nm$^{-1}$. 

Our observation of fixed pairings of RBM and C-mode frequencies indicates that we only observe chains for which the length does not affect the C-mode. This means that the chains consist of more than $100$ atoms (length $\geq 13\,$nm), in good agreement with X-ray diffraction measurements on bulk samples that reveal an average chain length of $40\,$nm~\cite{Shi:2016kqa}. We expect that Eqs.~\ref{EQ:CMODE_DIAMETER} and \ref{EQN:BG_DIAM} will no longer hold for short carbon chains since the C-mode frequency becomes dependent on the number of atoms for shorter chains. This does not exclude the existence of shorter chains in CNTs -- we merely expect them to occur at higher excitation energies and C-mode frequencies, i.e., as reported in Ref.~\cite{Zhao:2011co}. 

Experimentally, we observe no length dependence of the C-mode within our spectral and spatial resolution. Thus, the C-mode frequency does not carry information on the length of LLCCs beyond $100$ atoms. Long linear carbon chains inside double-walled carbon nanotubes can therefore be regarded as the finite realization of carbyne. It follows that the carbyne C-mode frequency and band gap predominantly depend on the local environment given by the chirality of the inner CNT. 

With few inner tube chiralities hosting LLCCs, the number of different local environments that affect the properties of the encapsulated chains is limited and distinct in nature. This fact is reflected in all experimental studies on linear carbon chains encapsulated in carbon nanotubes known to the authors~\cite{Zhao:2003dz, Fantini:2006ix,Jinno:2006gm,Nishide:2007bd,Shi:2011ey,Andrade:2015fka,Andrade:2015ke,Shi:2016kqa,Shi:2017cd}. The observed C-modes are well separated into different frequency bands that vary around comparable mean frequencies. In very good agreement with our measurements no C-modes between $1810\,$cm$^{-1}$ and $1825\,$cm$^{-1}$ are reported. According to Eq.~\ref{EQ:CMODE_DIAMETER}, only the C-mode of chains encapsulated in the metallic $(9,0)$ and $(8,2)$ nanotubes fall in this range. The fraction of metallic inner tubes with diameters below $1\,$nm, however, is very low, explaining why these C-mode frequencies are not observed experimentally~\cite{Pfeiffer:2003hg,Pfeiffer:2004kg,Pfeiffer:2005ho}.

Our results suggest that tunability of carbyne (C-mode frequency, electronic bandgap) can be achieved by choosing the right nanotube host, a property that can have interesting applications in low-dimensional optoelectronics. We show (i) which inner nanotube diameters and chiralities should generally be aimed for and (ii) which C-mode frequency should be used to monitor the growth of carbyne with the desired properties. 

In conclusion, using tip-enhanced Raman spectroscopy we have investigated isolated pairs of double-walled carbon nanotubes and encapsulated long linear carbon chains. The chain's Raman frequency is governed by the chirality of the inner tube and reduces with decreasing inner tube diameter. This points towards van-der-Waals forces as the dominating interaction mechanism between the host nanotube and the encapsulated chain. No length dependence of the chain's Raman mode frequency is evident for the long linear carbon chains investigated, suggesting that they can be viewed as a close to perfect representation of carbyne. Our experimental data establishes a firm link between the local environment of the host nanotube and properties of the encapsulated linear carbon chain, and identifies parameters for tuning the phononic and electronic properties of carbyne for device applications. 

\begin{acknowledgements}
The authors thank M.~Frimmer, M.~Parzefall and S.~Reich for fruitful discussions. This research was supported by the Swiss National Science Foundation (grant no. $200021\_165841$). S.~Heeg acknowledges financial support by ETH Z\"urich Career Seed Grant SEED-16 17-1.
\end{acknowledgements}


%

\pagebreak
\begin{widetext}
\textbf{In the following we present Supporting Information: Carbon nanotube chirality determines properties of encapsulated linear carbon chain}
\end{widetext}

\pagebreak
\begin{center}
\textbf{\large Supporting Information for Carbon nanotube chirality determines properties of encapsulated linear carbon chain}
\end{center}

Here we provide the following supporting information: topography recorded during TERS imaging; effect of encapsulated chains on the RBM of DWCNTs; assigning chiralities to the RBMs of inner carbon nanotubes; assigning chain band gaps to pairings of RBM and C-mode.

\section{Topography recorded during TERS imaging}
\begin{figure}[b]
\includegraphics[width=\linewidth]{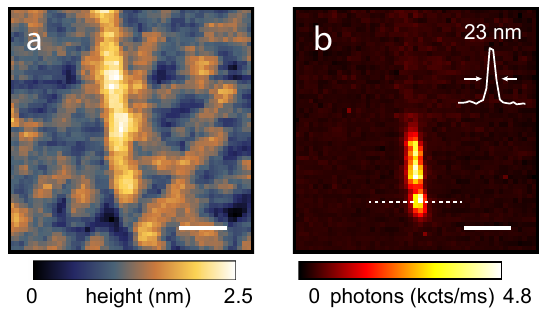}
  \caption{(a) AFM topography of double-walled carbon nanotube. (b) TERS image of linear carbon chain encapsulated in the DWCNT shown in (a). The scale bar is $100\,$nm.}
  \label{FIGSI:TERS}
 \end{figure}

In Fig.~\ref{FIGSI:TERS}(a) we present the AFM topography recorded during TERS imaging of the long linear carbon chain in Fig.~\ref{FIGSI:TERS}(b), which was also shown in Fig.~1 of the main paper. The TERS image shows that the lower part of the DWCNT in Fig.~\ref{FIGSI:TERS}(a) contains a LLCC. The topography and the TERS image are slightly offset, which is not uncommon in TERS experiments. The natural roughness of the substrate and contaminations of the sample surface generate the additional topographic features present in Fig.~\ref{FIGSI:TERS}(a), which partially leads to a modulation of the TERS signal due to variation in the tip-sample distance. 

\section{Effect of encapsulated chains on the RBM of DWCNT}
Here we show that the presence of encapsulated long linear carbon chains does not affect the RBM frequencies of the nanotubes forming a double-walled carbon nanotube. To achieve this we compare the Raman response of two double-walled nanotubes where both the inner and the outer CNTs have the same $(n,m)$ chiralities but only one of them (DWCNT\#1) hosts an encapsulated carbon chain. The other double-walled carbon nanotube (DWCNT\#2) is empty and serves as a reference. Both DWCNTs were measured in standard confocal configuration ($\lambda=633\,$nm, no tip) with an increased spectral resolution ($\sim1\,$cm$^{-1}$) compared to the TERS measurements in the main paper by using a high resolution grating ($1200\,$ grooves/mm instead of $600$ grooves/mm). DWCNT\#1 is one of the RBM/C-mode pairings at ($326\,$cm$^{-1}$/$1835\,$cm$^{-1}$) that was discussed and shown in Fig.~2 of the main paper.

\begin{figure}[b]
\includegraphics[width=\linewidth]{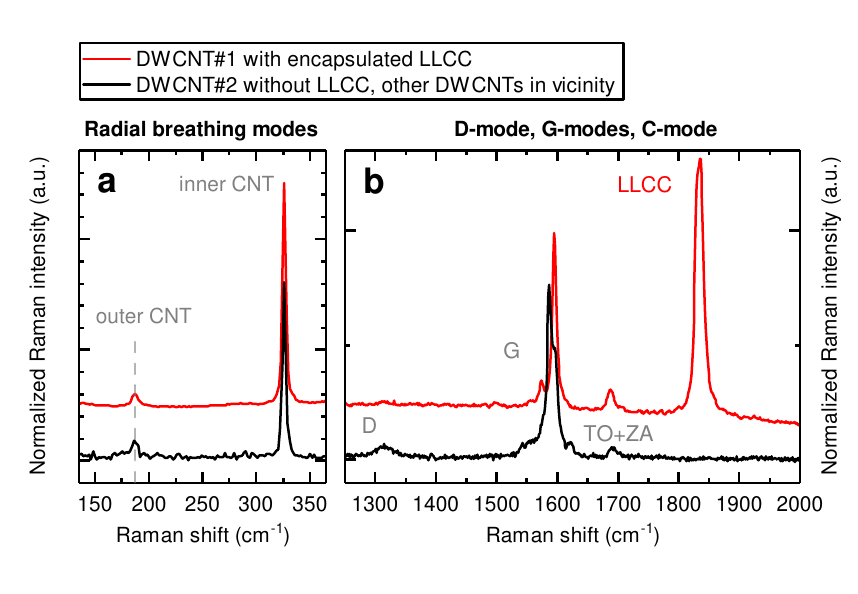}
  \caption{Raman spectra of two double-walled carbon nanotubes of the same chiralities, DWCNT\#1 with encapsulated long linear carbon chain (red) and DWCNT\#2 without chain (black). (a) Radial breathing modes of outer and inner CNTs. (b) D-mode and G-modes, and C-mode indicating the presence of an LLCC in DWCNT\#1. The spectra in (a) are individually normalized to comparable intensities. Spectra corresponding to DWCNT\#1 and DWCNT\#2, respectively, are offset for clarity.}
  \label{FIGSI:RBM_chain_nochain}
 \end{figure}

The Raman spectra of DWCNT\#1 (red) and DWCNT\#2 (black) are shown for the spectral region of the RBMs in Figure~\ref{FIGSI:RBM_chain_nochain}(a) and for the spectral region of the higher energy CNT modes (D, G, TO+ZA combination mode~\cite{Vierck:2017gk}) as well as the chain's C-mode in Fig.~\ref{FIGSI:RBM_chain_nochain}(b). The RBMs of both DWCNT were fitted using Lorentzian lineshapes and the results are listed in Table S1. The positions of the inner and outer RBMs, Fig.~\ref{FIGSI:RBM_chain_nochain}(a), are in excellent agreement for DWCNT\#1 and DWCNT\#2. Furthermore, the relative peak intensities of inner and outer RBMs are very similar for DWCNT\#1 and DWCNT\#2. In summary this unequivocally confirms that the chiralities of inner and outer tubes are indeed the same for both DWCNTs. Only DWCNT\#1, however, shows a C-mode peak associated with an encapsulated LLCC, Fig.~\ref{FIGSI:RBM_chain_nochain}(b), while no indication for the presence of a chain is found for DWCNT\#2. We conclude that encapsulated linear carbon chains do not change the RBM frequencies of their host tubes for double-walled carbon nanotubes. 

\begin{table}[t]
\label{TAB:RBMS} 
\begin{ruledtabular}
\begin{tabular}{c c c c }
& RBM outer CNT & RBM inner CNT & C-mode \\
DWCNT\#1 & $187.1\,$cm$^{-1}$ & $326.2\,$cm$^{-1}$ & $1834.4\,$cm$^{-1}$ \\
DWCNT\#2 & $186.6\,$cm$^{-1}$ & $326.1\,$cm$^{-1}$ & - \\
\end{tabular}
\end{ruledtabular}
\caption{RBM frequencies of inner and outer carbon nanotubes of DWCNT\#1 and DWCNT\#2 obtained from fitting the spectra in Fig.~\ref{FIGSI:RBM_chain_nochain} with Lorentzians. Only DWCNT\#1 contains a long-linear carbon chain as evidenced by the C-mode, while DWCNT\#2 is empty.}
\end{table}

This behaviour may seem surprising as the encapsulation of molecular structures in single-walled carbon nanotubes, i.e. J-aggregated molecular dyes, typically modifies the RBM frequencies by several cm\,$^{-1}$~\cite{Loi:2010dy,Gaufres:2013ej}. Double-walled carbon nanotubes, on the other hand, are far more rigid structures than SWCNTs because the inner and outer tubes mutually support and stabilize each other as observed in pressure-dependent Raman measurements~\cite{Arvanitidis:2005bs}. This explains why the presence of an encapsulated linear carbon chain does not change the experimentally observed RBM frequencies for double-walled carbon nanotubes. 

There are additional tubes close to DWCNT\#2 which contribute to the corresponding G-peaks in Fig.~\ref{FIGSI:RBM_chain_nochain}. For this reason it is not possible to extract information on doping of the nanotubes and charge transfer between chain and tubes through changes in the G-mode shape by comparing the G-modes of DWCNT\#1 and DWCNT\#2.  

\section{Assigning chiralities to the RBMs of inner carbon nanotubes}
Here we discuss the assignment of RBMs the chirality of the inner nanotubes for all parings of RBM and C-mode observed in the main paper. Unless stated otherwise, we do not consider carbon nanotubes chiralities whose optical transition energies are separated by more than $250\,$meV from the excitation energy $1.96\,$eV, as the observation of the corresponding RBM is highly unlikely even with the strong signal enhancement provided by TERS.

\subsection{RBM at $\mathbf{327\pm1\,}$cm$\mathbf{^{-1}}$ to (6,5) chirality} 
\begin{figure}[b]
\includegraphics[width=\linewidth]{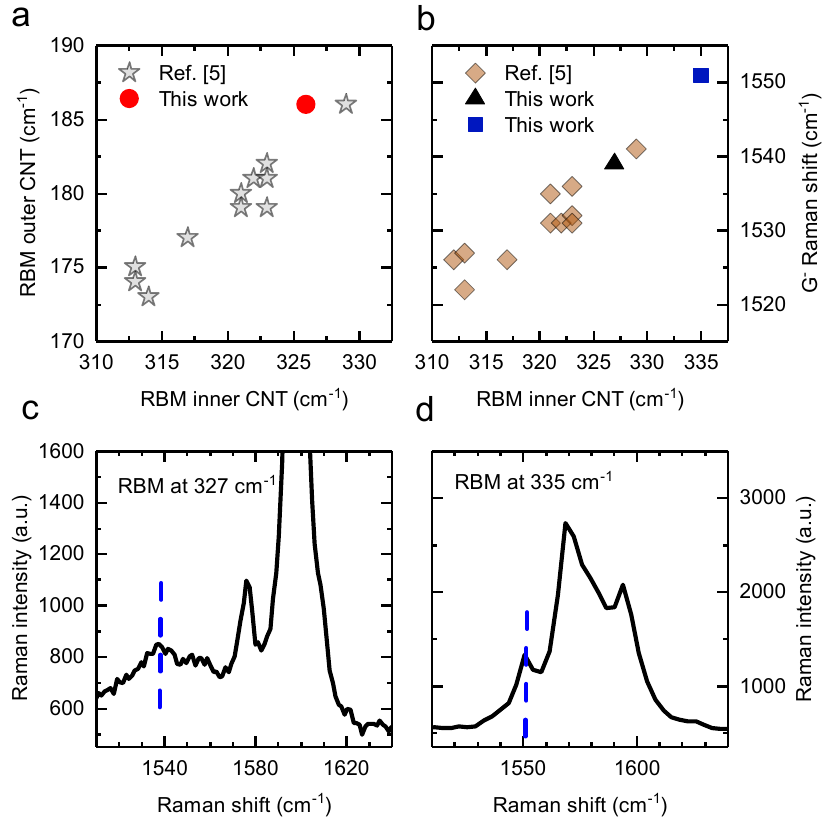}
  \caption{(a) Radial breathing modes of inner $(6,5)$ carbon nanotubes plotted against radial breathing modes of different outer nanotubes of metallic character. (b) Correlation between RBM and G$^{-}$-mode of inner $(6,5)$ carbon nanotubes. Partially adapted from Ref.~\cite{VillalpandoPaez:2010dd}. (c) G-mode Raman spectra of $(6,5)$ inner nanotube with RBM at $327\,$cm$^{-1}$. (d) G-mode Raman spectra of $(6,5)$ inner nanotube with RBM at $335\,$cm$^{-1}$.}
  \label{FIGSI:RBM_65}
 \end{figure}

The C-mode/RBM pairing with the inner tube's RBM at $327\pm1\,$cm$^{-1}$ is the only DWCNT where we also observe the RBM of the outer tube at $187\,$cm$^{-1}$. Its characteristics have already been discussed in the previous section, c.f. Fig.~\ref{FIGSI:RBM_chain_nochain} and Tab.~S1. A Raman study on isolated DWCNTs by \textit{Villalpando-Paez et al.} found very similar combinations of RBMs of inner and outer tubes~\cite{VillalpandoPaez:2010dd}. It was shown that the inner tube is the semiconducting $(6,5)$ chirality and the outer tubes are metallic with different diameters. By plotting inner vs. outer RBM, the authors of Ref.~\cite{VillalpandoPaez:2010dd} showed that the RBM of the $(6,5)$ tube increases with increasing (decreasing) RBM (diameter) of the outer metallic nanotube, see discussion in the main paper. We reproduce the data from Ref.~\cite{VillalpandoPaez:2010dd} as stars in Fig.~\ref{FIGSI:RBM_65}(a), and include our combination of inner and outer RBM as a red circle. It is in very good agreement with the correlation between inner and outer RBM observed in Ref.~\cite{VillalpandoPaez:2010dd} and confirms that the RBM at $327\pm1\,$ arises from the $(6,5)$ inner nanotube. 

\textit{Villalpando-Paez et al.} further reported that the lower G-mode component (G$^{-}$) of the $(6,5)$ inner tube also depends on the outer tube~\cite{VillalpandoPaez:2010dd}. Their data is reproduced as brown diamonds in Fig.~\ref{FIGSI:RBM_65}(b). We observe G$^{-}$ at $1539\pm1\,$cm$^{-1}$, spectrum see Fig.~\ref{FIGSI:RBM_65}(c), which we plot as the black triangle in Fig.~\ref{FIGSI:RBM_65}(b). It is in very good agreement with the trend reported in Ref.~\cite{VillalpandoPaez:2010dd} and provides further evidence that the observed RBM arise from the $(6,5)$ inner carbon nanotube. 

\subsection{RBM at $\mathbf{335\pm1\,}$cm$\mathbf{^{-1}}$ to (6,5) chirality}
Based on a Raman study on bulk quantities of DWCNTs with tunable excitation wavelengths, the RBM at $335\,$cm$^{-1}$ may arise from either the $(6,5)$ or the $(6,4)$ chirality. As we do not observe the RBM of the outer nanotube, we use the position of the G$^{-}$-mode to show that the RBM belongs to the $(6,5)$ chirality. 

The $G$-modes of the DWCNT with the RBM at $335\,$cm$^{-1}$ are shown in Fig.~\ref{FIGSI:RBM_65}(d). The G$^{-}$-mode associated with the inner CNT appears at $1551\,$cm$^{-1}$ as indicated by the dashed line. We include the combination of RBM and G$^{-}$-mode as the blue square in Fig.~\ref{FIGSI:RBM_65}(b) and find a very good agreement with the trend observed in Ref.~\cite{VillalpandoPaez:2010dd}. The G$^{-}$-mode occurs exactly at the frequency that we expect for the $(6,5)$ tube with RBM at $335\,$cm$^{-1}$, which provides strong evidence that the RBM indeed arises from the $(6,5)$ inner nanotubes chirality. 

In the following we show that the RBM at $335\pm1\,$cm$^{-1}$ does not arise from the $(6,4)$ inner tube. We note that the increase of the G$^{-}$-mode of the inner nanotube with decreasing diameter of the outer nanotube is not specific to the $(6,5)$. This behaviour is generally observed in DWCNTs both in experiments as well as in theoretical studies~\cite{Levshov:2015hn,Levshov:2017iaa,Popov:2018uv}.

Studies on single-walled carbon nanotubes showed that the intrinsic RBM frequency of the $(6,4)$ nanotube is at $333\,$cm$^{-1}$, which is $\approx30\,$cm$^{-1}$ higher than that of the $(6,5)$ nanotube. The intrinsic position of the G$^{-}$-mode, however, is the same for both single-walled carbon nanotube chiralities ($1527\,$cm$^{-1}$)\cite{Maultzsch:2005fp,Telg:2012ex}. If the RBM at $335\pm1\,$cm$^{-1}$ were to belong to the $(6,4)$ tube, the outer nanotube had to have a diameter difference close to twice the graphene interlayer distance ($0.34\,$nm) such that it does not modify the experimentally observed RBM frequency. Consequently, we would also expect the G$^{-}$-mode close to its intrinsic frequency ($1527\,$\cm). This stands in disagreement with the experimentally observed G$^{-}$-mode position at $1551\,$cm$^{-1}$. It follows that the RBM at $335\pm1\,$cm$^{-1}$ does not arise from the $(6,4)$ inner tube. 

\subsection{RBM at $\mathbf{348\pm2\,}$cm$\mathbf{^{-1}}$ to (6,4) chirality} 
This assignment is straight forward because several Raman experimental studies on DWCNTs assigned the RBM at $348\pm1\,$\cm  to the $(6,4)$ inner nanotube without ambiguity~\cite{Pfeiffer:2003hg,Pfeiffer:2004kg,Pfeiffer:2005ho,VillalpandoPaez:2010hf}. In Fig.~\ref{FIGSI:RBM_G_348_364} we show the G$^{-}$-mode of this DWCNT, which is used as a reference in the next section. 

\subsection{RBM at $\mathbf{364\,}$cm$\mathbf{^{-1}}$ to (6,4) chirality}
The RBM at $364\,$\cm can in principle be associated with either the $(6,4)$ or the $(7,2)$ chirality. Its RBM frequency, however, is exactly between highest RBM frequency reported for the $(6,4)$ nanotube at $361\,$\cm (Ref.~\cite{Pfeiffer:2005ho}) and the lowest RBMs reported for the $(7,2)$ nanotube in the range $366\,$\cm to $372\,$\cm (Refs.~\cite{Pfeiffer:2005ho,VillalpandoPaez:2010hf}). Similar to our approach for the RBM $335\,$\cm, we will use the G$^{-}$ mode to show that the RBM at $364\,$cm$^{-1}$ arises from the $(6,4)$ inner carbon nanotubes. 

\begin{figure}
   \includegraphics[width=\linewidth]{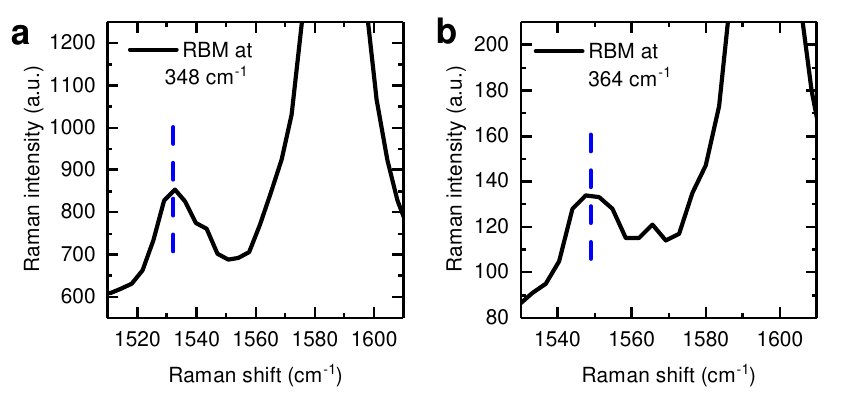}
  \caption{(a) G-mode TERS spectrum of DWCNT with RBM at $348\,$\cm. (b) G-mode TERS spectrum of DWCNT with RBM at $364\,$\cm. The dashed arrows indicates the position of the G$^{-}$-mode at $1532\,$\cm and $1549\,$\cm in (a) and (b), respectively.}
  \label{FIGSI:RBM_G_348_364}
\end{figure}

In Fig.~\ref{FIGSI:RBM_G_348_364}(a) we show G$^{-}$ mode spectra of the RBM at $348\,$\cm, which we assigned to the $(6,4)$ nanotube in the previous section, and indicate the position of the G$^{-}$-mode at $1532\,$\cm with the dashed line. Fig. ~\ref{FIGSI:RBM_G_348_364}(b) shows the corresponding spectrum for the RBM at $364\,$\cm, where we find the G$^{-}$ at $1549\,$\cm. Assuming that both RBM arise from the $(6,4)$ inner carbon nanotubes allows us to compare the peak positions directly. We observe an upshift of $16\,$\cm for the RBM and an upshift of $17\,$\cm for the G$^{-}$ mode. Such uniform increase in both RBM and G$^{-}$ frequency is exactly what we expect if both RBM arise from the same inner nanotube as we already discussed for the $(6,5)$ chirality, c.f. Fig.~\ref{FIGSI:RBM_65}(b) and Refs.~\cite{VillalpandoPaez:2010hf,Levshov:2015hn,Levshov:2017ia,Popov:2018uv}. Using the same arguments -- skipped for brevity --  it can be shown that the G$^{-}$ observed in Fig.~\ref{FIGSI:RBM_65}(b) mode cannot arise from the $(7,2)$ nanotube. We conclude that the RBM at $364\,$\cm arises from the $(6,4)$ inner nanotube. 

\subsection{RBM at $\mathbf{371\,}$cm$\mathbf{^{-1}}$ to (7,2) chirality} 
The RBM at $371\,$\cm can be unequivocally assigned to the $(7,2)$ nanotube by comparison with the data presented in Refs.~\cite{Pfeiffer:2005ho,VillalpandoPaez:2010hf}.

\subsection{RBM at $\mathbf{391\,}$cm$\mathbf{^{-1}}$ to (8,0) chirality} 
Here we show that the $(8,0)$ nanotube is the most likely inner nanotube candidate giving rise to the RBM at $391\,$\cm. \textit{Pfeiffer et al.} observed a RBM at $385\,$\cm with $1.83\,$eV excitation, which agreed well with their calculated RBM frequency of $382\,$\cm expected for the $(8,0)$ nanotube\cite{Pfeiffer:2003hg}. However, this peak was not assigned to a chirality, most likely due to its comparably low Raman intensity. 

We did not observe a G$^{-}$ mode in the relevant the frequency range below $1530\,$\cm~\cite{Telg:2012ex}. This speaks in favor of the $(8,0)$ chirality, because zigzag carbon nanotubes -- all chiralities with $(n,0)$ - only have one G-mode component with frequencies $\geq1580\,$\cm\cite{Thomsen:2007vc,Michel:2010cb,Hirschmann:2014kia}.

To determine wether we can expect to observe the RBM of the $(8,0)$ chirality with our excitation ($1.96\,$eV), we estimate the optical transition energy of the $(8,0)$ chirality as the inner nanotube of a DWCNT. For SWCNTs in solution, the transition energy of the $(8,0)$ chirality was reported at $1.97\,$eV with an RBM frequency of $361\,$\cm.  The $30\,$\cm upshift in RBM frequency is typically accompanied by a downshift in optical transition energy. As a crude approximation, we take the shift rate of $-2.8\,$meV/\cm  for the $(6,5)$ chirality reported in Ref.~\cite{Pfeiffer:2005ho} and arrive at an optical transition energy of $1.88\,$eV for the $(8,0)$ tube as the inner nanotube of a DWCNT. This is still reasonably close to our excitation and agrees well with the relatively strong intensity of the RBM that we observe experimentally even in confocal mode without TERS enhancement (not shown). In summary we conclude that the RBM at $391\,$\cm arises from the $(8,0) $ nanotube. 

\section{Assigning Chain Band Gaps to pairings of RBM and C-mode}
\begin{figure}[t]
\includegraphics[width=\linewidth]{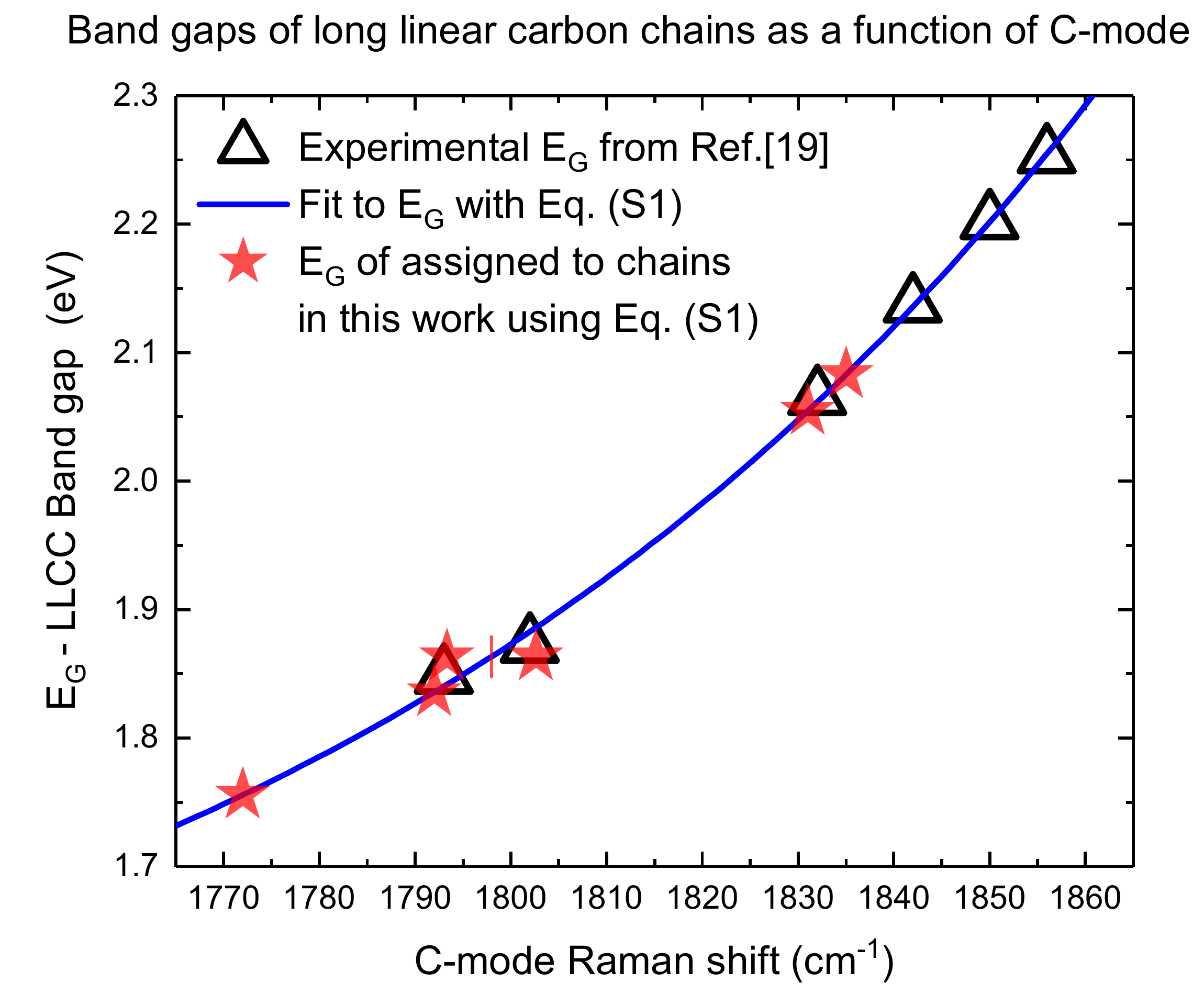}
  \caption{LLCC band gap E$_{\textrm{G}}$ as a function of C-mode (triangles) from Raman measurements on bulk quantities of LLCCs encapsulated in DWCNTs, c.f. Ref.~\cite{Shi:2017cd}. Red stars mark the band gaps extracted for the C-modes observed in the main paper according to Eq.~\ref{EQNSI:bandgaps}.}
  \label{FIGSI:BandGaps}
 \end{figure}

Here we show how we assign band gaps to the pairings of RBMs and C-mode frequency and subsequently to the inner tube's diameter as presented in Fig.~4(c) and in the text of the main paper. In a recent study Raman spectroscopy with tunable laser sources was performed on bulk quantities of the same DWCNT/LLCC starting materials used in this work~\cite{Shi:2017cd}. Through Raman resonance profiles it was shown that the C-mode frequency of a linear carbon chain is unequivocally connected to the chain's bandgap. We plot the C-mode frequency and corresponding band gaps $E_G$ that were experimentally obtain in Ref.~\cite{Shi:2017cd} as black triangles in Fig.~\ref{FIGSI:BandGaps}, c.f. Fig~4(c) of the main paper. We found the best fit to the data presented in Ref.~\cite{Shi:2017cd} to be of exponential form such that the band gap $E_G$ of LLCCs as a function of C-mode frequency is given by

\begin{equation}\label{EQNSI:bandgaps}
E_{\textrm{gap}}(\omega_C)=1.44\,\textrm{eV} + 0.263\,\textrm{eV} \exp\left( \frac{\omega_C-1756\,\textrm{cm}^{-1}}{{88.8\, \textrm{cm}^{-1}}} \right).
\end{equation}
Eq.~\ref{EQNSI:bandgaps} is plotted as a blue curve in Fig.~\ref{FIGSI:BandGaps} and in Fig~4(c) of the main paper.  It excellently describes the correlation between band gaps and the corresponding C-mode frequencies as obtained in Ref.~\cite{Shi:2017cd}. 

We make use of Eq.~\ref{EQNSI:bandgaps} to assign bandgaps to the C-mode frequencies measured by TERS in the main paper, and plot them as red stars in Fig.~\ref{FIGSI:BandGaps}, c.f. Fig~4(c) of the main paper. Note that all except one C-mode frequency are very similar to the ones observed in bulk measurements in Ref.~\cite{Shi:2017cd}. 

Relation (1) in the main paper established a correlation between the diameter of the inner carbon nanotubes and the C-mode frequency of the encapsulated long linear carbon chain. As the C-mode frequency is correlated with the band gap of the chain, c.f. Fig.~\ref{FIGSI:BandGaps} and Eq.~\ref{EQNSI:bandgaps}, we are able to provide relation (2) of the main paper, which connects the diameter of the inner CNT with the bandgap of the chains.

\end{document}